\documentclass[aps,prl,twocolumn,groupedaddress]{revtex4}
\usepackage{graphicx}
\begin{document}
%\preprint{}


\title{Carrier-Concentration Dependence of the Pseudogap Ground State of Superconducting Bi$_2$Sr$_{2-x}$La$_x$CuO$_{6+\delta}$ Revealed by $^{63,65}$Cu-Nuclear Magnetic Resonance in Very High Magnetic Fields}

\author{Shinji Kawasaki$^1$}
\author{Chengtian Lin$^2$}
\author{Philip L. Kuhns$^3$}
\author{Arneil P. Reyes$^3$}
\author{Guo-qing Zheng$^{1,4}$}

\affiliation{$^1$Department of Physics, Okayama University, Okayama 700-8530, Japan}
\affiliation{$^2$Max-Planck-Institut fur Festkorperforschung, Heisenbergstrasse 1, D-70569 Stuttgart, Germany}
\affiliation{$^3$National High Magnetic Field Laboratory, Tallahassee, Florida 32310, USA}
\affiliation{$^4$Institute of Physics and Beijing National Laboratory for Condensed Matter Physics, Chinese Academy of Sciences, Beijing 100190, China}


%\author{}
%\thanks{}
%\affiliation{}


%\date{\today}

\begin{abstract}
 We report the results of the Knight shift by $^{63,65}$Cu-nuclear-magnetic resonance (NMR) measurements on single-layered copper-oxide Bi$_2$Sr$_{2-x}$La$_x$CuO$_{6+\delta}$ conducted under very high magnetic fields up to 44 T.  The magnetic field suppresses superconductivity completely and the pseudogap ground state is  revealed. The  $^{63}$Cu-NMR Knight shift shows that there remains a finite  density of states (DOS) at the Fermi level in the zero-temperature limit, which indicates that the pseudogap ground state is a metallic state with a finite volume of Fermi surface. The residual DOS in the pseudogap ground state decreases with decreasing doping (increasing $x$) but remains quite large even at the vicinity of the magnetically ordered phase of $x\ge$  0.8, which suggests that the DOS plunges to zero upon approaching the Mott insulating phase.
\end{abstract} 

%\pacs{}
%\keywords{}

\maketitle


%************************intruduction*************************************************************************************
The mechanism of the high transition temperature ($T_c$) superconductivity in copper oxides (cuprates) \cite{Bednorz} still remains unclear, largely because the relationship between the normal-state properties and superconductivity is unclear. In conventional metals, superconductivity develops out of a Fermi liquid state. This is also true in the electron-doped $n$-type cuprates\cite{SatoPRL}. However, in the normal state of $p$-type cuprates, there is an intriguing phenomenon called a pseudogap state, in which the density of states (DOS) is depleted upon decreasing temperature ($T$) below a characteristic temperature $T^*$\cite{Timusk}. Several measurements have suggested that the pseudogap and superconductivity are coexisting states of matter\cite{ZhengBi2201,Tanaka,Yoshida}, but the detailed properties and the origin of the pseudogap are still under debate\cite{Moriya,Imada,Lee,ZhangRice}. Previous measurements have suggested that the pseudogap is either associated with disconnected Fermi-arcs\cite{Tanaka,Yoshida}, or small Fermi pockets\cite{Leyraud,Bangura}, or associated with coexisting Fermi arc and small Fermi pockets\cite{XJZhou}. In contrast, there was also a proposal that the Fermi surface shrinks to a nodal point when cooled to $T$ = 0\cite{Kanigel}. More importantly, it is still unknown how the pseudogap ground state (PGS) would evolve as the Mott insulating state is approached.

Experimentally, this is difficult due to the high upper critical field ($H_{c2}$ $\sim$ 100 T). In addition, there have been few systems which allow us to explore in a wide doping region. Bi$_2$Sr$_{2-x}$La$_{x}$CuO$_{6+\delta}$ is one of the ideal systems to study the subject. It has a single CuO$_2$ layer, highly two- dimensional structure\cite{Wang}, and much lower $T_c^{max}$ $\sim$ 32 K compared to other cuprates. Here, La introduces one electron to Bi$_2$Sr$_{2}$CuO$_{6+\delta}$\cite{Akimitsu}, hence $``1-x"$ corresponds to the hole doping rate. 

In this Letter, we report the results of the spin susceptibility via the $^{63,65}$Cu-NMR Knight shift measurements in Bi$_2$Sr$_{2-x}$La$_{x}$CuO$_{6+\delta}$ carried out under very high magnetic fields up to 44 T. When the quantity $1/T_1T$ has a strong $T$-dependence\cite{ZhengBi2201}, where $T_1$ is the spin-lattice relaxation time, it is difficult to extract the DOS from this quantity. In contrast, the Knight shift is directly proportional to the DOS, which allows us to evaluate the doping dependence of the residual DOS in the PGS. We find, for the first time, that an antiferromagnetic order occurs in the strongly underdoped region of this family of compounds. Most importantly, we find that the PGS is a metallic state with a finite DOS which decreases with decreasing doping but remains quite large even at the vicinity of the magnetically ordered phase. Our result suggests that the DOS  plunges to zero upon approaching the Mott insulating phase.

 Single crystalline Bi$_2$Sr$_{2-x}$La$_{x}$CuO$_{6+\delta}$ samples (0.0 $\le$ $x$ $\le$ 0.9) are grown by the traveling solvent floating zone method as reported elsewhere\cite{Lin1,Ando}. Black and shiny single crystal platelets sized up to $\sim$ 10 $\times$ 3 $\times$ 0.4 mm$^3$ cleaved from the as-grown ingot were used. 
For all measurements, the magnetic field is applied along the $c$-axis. High magnetic fields are generated by the Bitter magnet (21.7 - 30 T) and the Hybrid magnet (44 T), respectively, in the National High Magnetic Field Laboratory, Tallahassee, Florida. 
 The $T$ dependence of Cu-NMR/nuclear quadrupole resonance (NQR) intensity $I(0)$ is precisely estimated through a fitting to $I(t)=I(0)\exp{(-t/T_2)}$, where $T_2$ is the nuclear spin-spin relaxation time.

\begin{figure}[h]
\includegraphics[width=7.5cm]{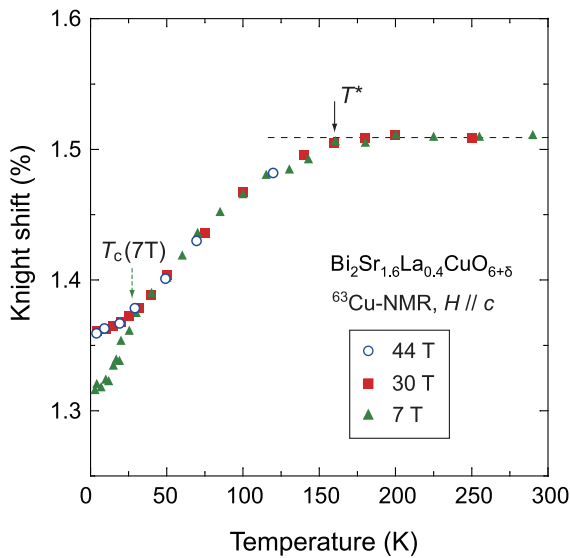}% Here is how to import EPS art
\caption{\label{fig:t1t} (Color online) $T$ dependence of $^{63}$Cu Knight shift ($H$ $\parallel$ $c$) for $x$ = 0.40 measured under different magnetic fields. The solid and dotted arrows indicate the pseudogap temperature $T^*$ and superconducting $T_c$ ($H$ = 7 T), respectively. The dotted line is guide to the eye.}
\end{figure}

Figure 1 shows the $T$ dependence of the Knight shift ($K_c(T)$) for the $x$ = 0.40 sample measured at $H$ = 7, 30, and 44 T, respectively. Here, the obtained $K_c(T)$ is written as $K_c(T) = K_s(T) + K_{orb}$, where $K_{\rm s}$ and $K_{orb}$ are the shifts due to the spin and the orbital susceptibility, respectively. Generally, $K_{orb}$ is $T$- and $H$-independent.  Most importantly, $K_{s}$ and $K_{orb}$ are expressed as $K_s(T) = A_{hf}^s \chi_s(T)$ and $K_{orb} = A_{hf}^{orb} \chi_{orb}$, where $A_{hf}$ is the hyperfine coupling constant.

At high $T$ region, $K_c$ does not depend on $T$, which is consistent with the observation of a full Fermi-surface above $T^*$\cite{Norman}. As indicated in the figure, $K_c$ starts to decrease below $T^*$ $\sim$ 160 K.  Notably, 1/$T_1T$ for $x$ = 0.40 also starts to decrease below $T^*$\cite{ZhengBi2201}. These results indicate a certain loss of the DOS at the Fermi surface taking place below $T^*$, i.e. the opening of a pseudogap. This $T$ dependence of $K_c$ is consistent with other high-$T_c$ cuprates\cite{Asayama}. As seen in the figure, $T^*$ is almost field independent. At $H$ = 7 T, $K_c$ decreases abruptly below $T_c(H)$ $\sim$ 30 K due to the reduction of the spin susceptibility as a result of a spin-singlet Cooper pairing. However, at $H$ = 30 and 44 T, no signature of superconducting transition is observed, as in previous $T_1$ measurement\cite{ZhengBi2201}. Also, the value of $K_c$ at the lowest $T$ is $H$-independent between $H$ = 30 and 44 T. This indicates that the superconductivity is suppressed by fields $H$ $\ge$ 30 T and that the $T$-dependence of $K_c$ ($H$ $\ge$ 30 T) represents the property of the PGS. Importantly, $K_c$ is quite large in the $T$ = 0 limit. Namely, the PGS has a finite residual DOS. Furthermore, no internal magnetic field is found when superconductivity is destroyed.

\begin{figure}[h]
\includegraphics[width=7.5cm]{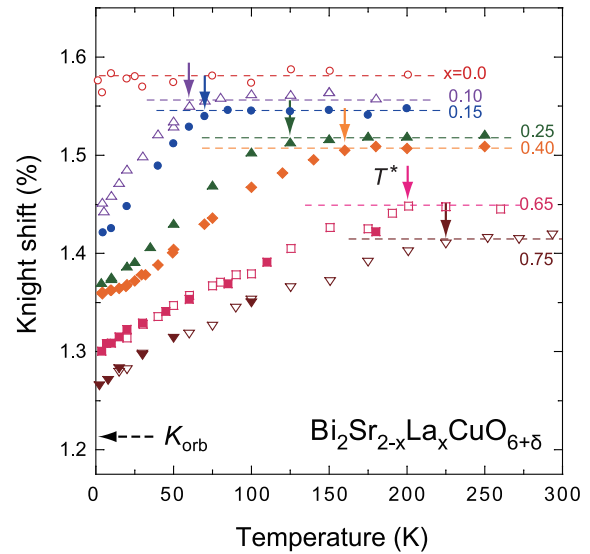}% Here is how to import EPS art
\caption{\label{fig:t1t} (Color online) $T$ dependence of $K_c$ for 0.0 $\le$  $x$ $\le$ 0.75. The open symbols are results measured at 9 T. The closed symbols are taken at different fields as follows: 30 T (for all $x$ $\ge$  0.15), 21.7 T (for $x$ = 0.65 and 0.75), and 44 T (for $x$ = 0.40). The solid arrows indicate the pseudogap temperature $T^*$. The dotted lines are guides to the eye. The dotted arrow indicates the value of $K_{orb}$. (see text)}
\end{figure}

We study the doping evolution of the residual DOS by performing the same measurements  for other samples with different $x$. Fig. 2 shows the summary of $T$ dependence of $K_c(T)$ for 0.0 $\leq$ $x$ $\leq$ 0.75 after suppressing superconductivity by the magnetic field. At high $T$, the constant value of $K_c$ increases systematically with increasing hole-doping (decreasing La content). As shown by arrows in the figure, $T^*$ can be clearly identified, below which $K_c$ starts to decrease, for 0.10 $\leq x \leq$ 0.75.

In order to quantitatively evaluate the residual DOS in the PGS, one needs to estimate $K_{orb}$, which we find to be 1.21\% as elaborated below. 
 In the superconducting state,  $K_c = K_{s} + K_{orb} + K_{dia}$. $K_{dia} = -H_{dia}/H$ is the contribution due to the diamagnetism in the vortex state, which is estimated  by using the relation $H_{dia}$ = ($\phi_0$/4$\pi$$\lambda_{ab}$$\lambda_{c}$)ln($\beta$$e^{-1/2}$$d$/$\sqrt{\xi_{ab}\xi_{c}}$)\cite{deGannes}, where $\phi_0$ is the flux quantum, $\lambda$ is the penetration depth, $d$ is the vortex distance, and $\beta$ = 0.381\cite{deGannes}. We have used $\lambda_{ab}$ = 4000 $\r{A}$, $\lambda_{c}$ = 10$\lambda_{ab}$ \cite{Russo}, $\xi_{ab}$ = 35 $\r{A}$, and the relation $\lambda_{c}$/$\lambda_{ab}$ = $\xi_{ab}$/$\xi_{c}$. The value of $\xi_{ab}$ is extracted from $H_{c2}$($H \| c$) $\sim$ 26 T obtained in the present work.

 Figure 3 shows the field dependence of $K_c$ for $x$ = 0.40 at $T$ $\sim$ 2 K.
  $K_s$ at $T << T_c$ in the superconducting state contains the contribution from nodal quasiparticles which is $H$-dependent and that due to impurity scattering\cite{Maki}. The former in the absence of impurity is proportional to $\sqrt{H}$, namely, $K_c(H)$ $-$ $K_{dia}$ = $K_{orb}$ $+$ $K_H$$\sqrt{H/H_{c2}}$\cite{Volovik} which is shown as dotted curve in Fig. 3.  The latter is not negligible for present sample, since 1/$T_1T$ well below $T_c$ remains finite even at $H$ = 0\cite{ZhengBi2201}. To account for the both effects in the dirty limit, we employed the theoretical model, $K_c(H)$ $\propto$ $\frac{H}{H_{c2}}$ln($\frac{H_0}{H}$)\cite{Vekhter}, where $H_0$ is a cutoff energy\cite{Vekhter}. This model has succeeded in reproducing the field dependence of $K_c$ for TlSr$_2$CaCu$_2$O$_{6.8}$\cite{ZhengTl}.  As shown by the solid curve in Fig. 3, the relation $K_c$ $-$ $K_{dia}$ = 1.29\% + 0.0261 ($\frac{H}{26.3}$ln($\frac{338}{H}$)) can reproduce experimental results reasonably well. The value of 1.29 \% is the sum of $K_{orb}$ and the impurity-induced spin shift, $K_s^{imp}$.  From the relation 1.29\% = $K_{orb}$ + ($\delta n_{imp}/n_0$)(1.38\% $-$ $K_{orb}$), we obtained the value of $K_{orb}$ = 1.21 \%. Here, $\delta n_{imp}/n_0$ = 0.468 is the value extracted from $T$ dependence of $1/T_1T$ at $H$ = 0\cite{ZhengBi2201} using theoretical calculation\cite{Bulut,ZhengTl}, and 1.38 \% is the value of $K_c$ at $T_c$ = 32 K. The obtained value of $K_{orb}$ = 1.21 \% is comparable to other high $T_c$ cuprates\cite{Asayama}. Thus, we can extract the residual DOS, $N_{res}(E_F) \propto K_s(T=0) = K_c(T=0) - K_{orb}$ from the results in Fig. 2. The results will be discussed later.

\begin{figure}[h]
\includegraphics[width=7.5cm]{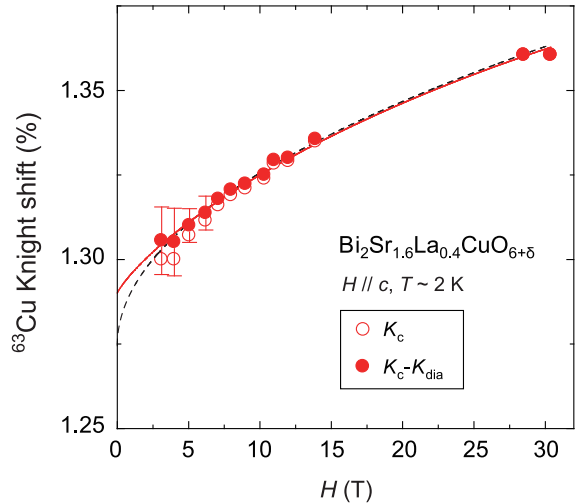}% Here is how to import EPS art
\caption{\label{fig:t1t} (Color online) Field dependence of $K_c$ and $K_c$ $-$ $K_{dia}$ for $x$ = 0.40. The dotted and solid curves are fittings assuming $K_c(H)$ $-$ $K_{dia}$ $\propto$  $\sqrt{H}$ in the absence of impurity scattering, and $K_c(H)$ $-$ $K_{dia}$ $\propto$ $\frac{H}{H_{c2}}$ln$(\frac{H_0}{H})$ in the presence of impurity scattering, respectively.}
\end{figure}

\begin{figure}[h]
\includegraphics[width=7.5cm]{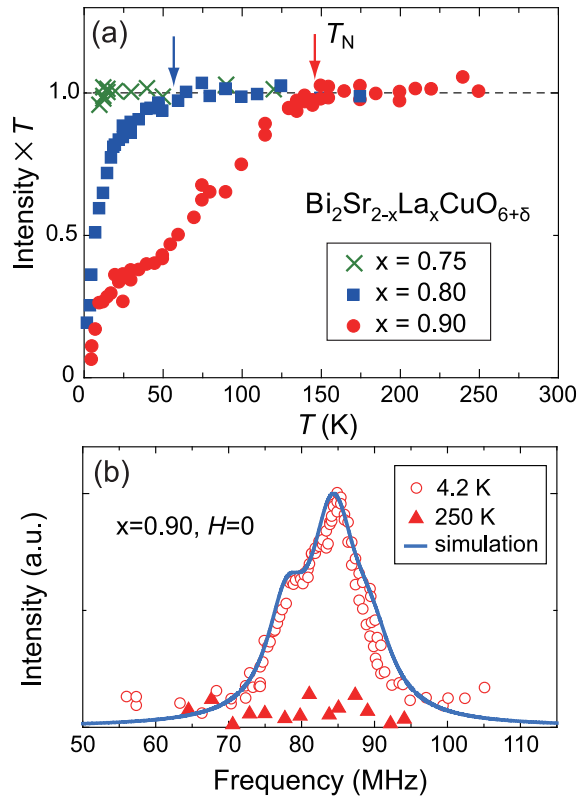}% Here is how to import EPS art
\caption{\label{fig:t1t} (Color online) (a)$T$ dependences of $I(T) \times$ $T$ for $x$ = 0.75, 0.80, and 0.90, respectively. The data are normalized by the value at high temperatures. The dotted line is a guide to the eye. The arrows indicate $T_N$.  (b) Zero-field $^{63,65}$Cu-NMR spectrum above and below $T_N$ for $x$ = 0.90. The solid curve is a fit to a model described in the text.}  
\end{figure}

Meanwhile, we turn to the strongly underdoped samples, $x$ = 0.90 and 0.80 which do not show superconductivity. Fig. 4(a) shows the $T$ dependences of $^{63,65}$Cu-NMR/NQR intensity times $T$. Generally, $I(T)$ increases in proportion to 1/$T$. Therefore, the quantity $I(T) \times$ $T$ must be constant. Actually, it is $T$ independent down to $T_c(H)$ for $x$ = 0.75. However, it decreases upon cooling below $T$ = 60 and 140 K for $x$ = 0.80 and 0.90, respectively. In addition, as seen in Fig. 4(b), we further found a zero-field NMR spectrum at low $T$ induced by the internal magnetic field ($H_{int}$) at Cu site. These results evidence the occurrence of antiferromagnetic order with $T_N$ = 60 K for $x$ = 0.80 and $T_N$ = 140 K for $x$ = 0.90. The spectrum can be modelled with a quadrupole-perturbed Zeeman transition with parameters:  $H_{int}$ = 7.2 T, $^{63}$$\nu_Q$($^{65}$$\nu_Q$) = 31.7(29.3) MHz, and $\theta$ = 57$^{\circ}$. Here, nuclear spin Hamiltonian is given as $\mathcal{H}_{AFM} = -\gamma\hbar\vec{I}\cdot\vec{H}_{int} + (h \nu_{Q}/6)[3{I_z}^2-I(I+1)]$ and $\theta$ is the angle between $\nu_Q$ and $H_{int}$. These parameters are in good agreement with previous results in insulating phase of Bi$_2$Sr$_2$CuO$_6$\cite{Kato}. It is noted that the system with $x$ $\ge$ 0.8 is in the insulating state as evidenced by resistivity measurements\cite{Ando}.

\begin{figure}[h]
\includegraphics[width=7.5cm]{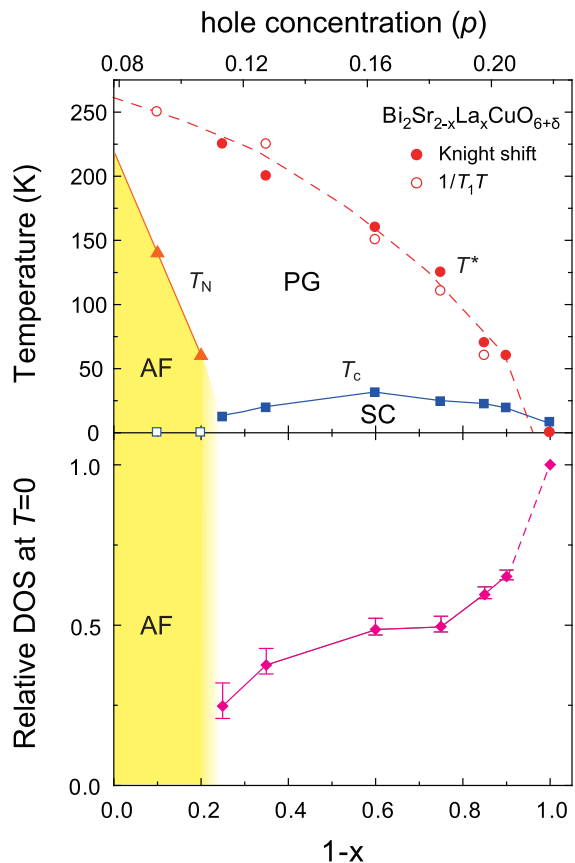}% Here is how to import EPS art
\caption{\label{fig:t1t} (Color online) Upper panel shows characteristic temperatures vs. 1$-$$x$ ($\propto$ hole concentration). For $x$ = 0.90 and 0.80, no superconductivity occurs (open squares). The open circles are taken from Ref\cite{ZhengBi2201,ShinjiAFM}. Lower panel shows the hole-concentration dependence of relative DOS at $T$ = 0.}
\end{figure}

Now we discuss the main finding. The upper panel of Fig. 5 shows the phase diagram  obtained through the present work together with $T^*$ determined by our  $T_1$ measurements\cite{ZhengBi2201}. $T^*$ for $x$ = 0.90 is determined by $T_1$ measurement\cite{ShinjiAFM}. The hole concentration in Bi$_2$Sr$_{2-x}$La$_x$CuO$_{6+\delta}$ was evaluated by Ando $et$ $al$ \cite{Ando} from Hall coefficient (upper horizontal scale), and has been used quite widely\cite{Wang2}.  From the $T$ dependence of $K_s(T)$, we obtained doping dependence of $T^*$.  At $x$ = 0, a metallic state is established\cite{ZhengBi2201}. From $x$ = 0.10 to 0.90, notably, $K_s(T)$ and 1/$T_1T$ indicate almost the same $T^*$.  For $x$ = 0.80 and 0.90, antiferromagnetic order sets in although $T^*$ still exists at $T^*$ = 250 K for $x$ = 0.90.

The lower panel of Fig. 5 shows the evolution of residual DOS for the PGS. The vertical axis is the relative DOS, $N_{res}$/$N_0$, at $T$ = 0 as defined by $N_{res}$/$N_0$ $=$ $K_s(T=0)/K_s(T=T^*)$. For $x$ $\textgreater$ 0, this quantity contains the contribution released by suppressing superconductivity, and the contribution by impurity scattering which does not change when the superconductivity is suppressed. The latter contribution is about 0.1\% and is almost $x$-independent. For $x$ = 0, the system is in a conventional metallic state, which remains true when the superconductivity is suppressed.  For $x$ $\textgreater$ 0.1, a pseudogap opens and causes the partial loss of the DOS. The relative DOS continuously decreases with decreasing hole concentration. It is interesting that the relative DOS remains quite large ($\sim$ 25 \%) even at $x$ = 0.75, which is very close to the Mott state at $x$ = 0.80.

The result suggests that the residual DOS for the PGS will vanish abruptly as the Mott insulating phase is approached. 
Recent experiment suggests that a symmetry breaking may occur in the pseudogap state\cite{Hashimoto}. Our result puts a constraint on the searching for such symmetry breaking. Namely, any symmetry breaking, if it would occur, should not be accompanied by an appearance of an internal magnetic field. Furthermore, the relative DOS we found in this experiment ($\sim$ 25 \%) is much larger than that inferred from the quantum oscillation ($\sim$ 3 \%)\cite{Leyraud}.

In conclusion, NMR measurements under very high magnetic fields  up to 44 T reveal the antiferromagnetically ordered state in the strongly underdoped region and the doping evolution of pseudogap ground state in Bi$_2$Sr$_{2-x}$La$_{x}$CuO$_{6+\delta}$. 
Our result indicates that the pseudogap ground state is a metallic state with a finite DOS which decreases with decreasing doping but remains quite large even at the vicinity of the Mott insulating phase. The present work revealed the complete phase diagram for a single-layered high-$T_c$ cuprate.

We thank P. A. Lee, T. Yokoya, M. Imada, Y. Fuseya, H. Kohno, K. Miyake, and X. G. Wen  for helpful comments, and R. Smith, K. Shimada, and T. Tabuchi for their experimental help. This work was supported in part by research grants from MEXT (No.17072005, No. 21102514, and No. 22103004), JSPS (No. 20244058). A portion of this work was performed at the National High Magnetic Field Laboratory, which is supported by NSF, the State of Florida, and DOE.
%\bibliography{apssamp}

\end{document}